# Variance-based sensitivity analysis: The quest for better estimators and designs between explorativity and economy

Samuele Lo Piano[a,#], Federico Ferretti[b], Arnald Puy[c,d], Daniel Albrecht[e], and Andrea Saltelli[f]

[a] School of the Built Environment, University of Reading (United Kingdom)

[b] European Commission, Joint Research Centre, 'Energy Security, Distribution & Markets' Unit, Ispra (Italy)

[c] Department of Ecology and Evolutionary Biology, Princeton University, Princeton, New Jersey (USA)

[d] Centre for the Study of the Sciences and the Humanities, University of Bergen (Norway)

[e] European Commission, Joint Research Centre - JRC-I.2, Competence Center on Modelling, Ispra (Italy)

[f] Open Evidence Research,
Universitat Oberta de Catalunya (UOC), Barcelona (Spain)

[#] Corresponding Author, s.lopiano@reading.ac.uk

## Abstract

Variance-based sensitivity indices have established themselves as a reference among practitioners of sensitivity analysis of model outputs. A variance-based sensitivity analysis typically produces the first-order sensitivity indices $S_j$ and the so-called total-effect sensitivity indices $T_j$ for the uncertain factors of the mathematical model under analysis.

Computational cost is critical in sensitivity analysis. This cost depends upon the number of model evaluations needed to obtain stable and accurate values of the estimates. While efficient estimation procedures are available for $S_j$ (Tarantola et al., 2006), this availability is less the case for $T_j$ (Iooss and Lemaître, 2015). When estimating these indices, one can either use a sample-based approach whose computational cost depends on the number of factors or use approaches based on meta-modelling/emulators (e.g., Gaussian processes).

The present work focuses on sample-based estimation procedures for $T_j$ for independent inputs and tests different avenues to achieve an algorithmic improvement over the existing best practices. To improve the exploration of the space of the input factors (design) and the formula to compute the indices (estimator), we propose strategies based on the concepts of *economy* and *explorativity*. We then discuss how several existing estimators perform along these characteristics.

Numerical results are presented for a set of seven test functions corresponding to different settings (few important factors with low cross-factor interactions, all factors equally important with low cross-factor interactions, and all factors equally important with high cross-factor interactions). We conclude the following from these experiments: a) sample-based approaches based on the use of multiple matrices to enhance the economy are outperformed by designs using fewer matrices but with better explorativity; b) among the latter, asymmetric designs perform the best and outperform symmetric designs having corrective terms for spurious correlations; c) improving on the existing best practices is fraught with difficulties; and d) ameliorating the results comes at the cost of introducing extra design parameters.

***Table 1** - Legend*

| | |
|---|---|
| $\boldsymbol{A}, \boldsymbol{B}, \boldsymbol{C}, \dots$ | Sample matrices |
| $\boldsymbol{A}_{\boldsymbol{B}}^{(j)}$ | Sample matrix where all columns are from $\boldsymbol{A}$ except for column $j$, which is from $\boldsymbol{B}$; likewise for other sample matrices $\boldsymbol{C}, \boldsymbol{D}, \boldsymbol{E}$ and so forth |
| $\boldsymbol{a}_i, \boldsymbol{b}_i, \boldsymbol{c}_i, \dots$ | $i^{th}$ row of matrices $\boldsymbol{A}, \boldsymbol{B}, \boldsymbol{C}$, etc., respectively |
| $\boldsymbol{a}_{\boldsymbol{b}i}^{(j)}$ | $i^{th}$ row of matrix $\boldsymbol{A}_{\boldsymbol{B}}^{(j)}$ |
| $e$ | Economy of a given design, defined as the number of elementary effects useful to compute $T_j$ (defined as $E_T/N_T$) |
| $ee$ | Generic elementary effect |
| $E_T$ | Total number of elementary effects |
| $E_{X_j}(\cdot), V_{X_j}(\cdot)$ | Expected value and variance of argument $(\cdot)$ taken over factor $X_j$ |
| $E_{\boldsymbol{X}_{\sim j}}(\cdot), V_{\boldsymbol{X}_{\sim j}}(\cdot)$ | Expected value and variance of argument $(\cdot)$ taken over all factors but $X_j$ |
| $H$ | Generic sample matrix $\boldsymbol{A}, \boldsymbol{B}, \boldsymbol{C}, \dots \boldsymbol{Z}$ |
| $i$ | Running index for the rows of a sample matrix $i = 1,2, \dots N$ |
| $j$ | Running index for factor $j = 1,2, \dots k$ |
| $k$ | Number of factors |
| $l$ | Running index over factor $j$; $l = 1,2, \dots j$ |
| $m, q$ | Running indices for the pool of sample-matrices (e.g., $H_1 = A$, and $H_{1-2}^{(j)} = A_B^{(j)}$) |
| $n$ | Number of sample matrices |
| $N$ | Column-dimension (length) of a single sample matrix |
| $N_T$ | Total number of points in the design |
| $p$ | Running index for the block on which the algorithm is executed (each block has column length $N = 2^p$) |
| $r$ | Running index for the repetition $r = 1,2, \dots 50$ |
| $s$ | Running index for the block with a power of two $s = 1,2, \dots k-1$ |
| $S_j$ | First-order effect sensitivity index for a generic factor $j$ |
| $T_j$ | Total-effect sensitivity index for a generic factor $j$ |
| $\chi$ | Explorativity, the fraction of non-repeated coordinates in the design |

# 1. Introduction

The sensitivity analysis of mathematical models aims to 'apportion the output uncertainty to the uncertainty in the input factors' (Saltelli and Sobol', 1995). Uses of sensitivity analysis are found in quality assurance, model calibration, model validation, uncertainty reduction, and model simplification, which are just a few among the possible applications.

Over the last three decades, sensitivity analysis (SA) has made steps to establish itself as a self-standing discipline with a community of practitioners gathering around the SAMO (Sensitivity Analysis of Modelling Output) international conferences since 1995. Special issues have been devoted to SA (Borgonovo and Tarantola, 2012; Ginsbourger et al., 2015; Helton et al., 2006; Saltelli, 2009; Tarantola and Saint-Geours, 2015; Tarantola and Saltelli, 2003; Turányi, 2008), mostly in relation to the SAMO events. Available textbooks for sensitivity analysis include Borgonovo (2017), Cacuci (2003), de Rocquigny et al. (2008), Fang et al. (2005), and Saltelli et al. (2008, 2004, 2000). SA is acknowledged as a useful practice in model development and applications. Its use in regulatory settings (e.g., in impact assessment studies) is prescribed in guidelines both in Europe and the United States (European Commission, 2015; Office of Management and Budget, 2006; US EPA, 2015). SA is also an ingredient of sensitivity auditing (Saltelli et al., 2013; Saltelli and Funtowicz, 2014), a procedure to investigate the relevance and plausibility of model-based inference as an input to policy (European Commission, 2015; Science Advice for Policy by European Academies, 2019).

Tools such as sensitivity analysis and sensitivity auditing are particularly needed at this point in time when the accuracy, relevance and plausibility of the statistical and mathematical models used to support policy are often the subject of controversy (Jakeman et al., 2006; Padilla et al., 2018; Pilkey and Pilkey-Jarvis, 2007; Saltelli and Funtowicz, 2017; Saltelli and Giampietro, 2017), including at the time of submitting the present article, the COVID-19 pandemic (Saltelli et al., 2020; Steinmann et al., 2020). As highlighted elsewhere (Lo Piano and Robinson, 2019; Saltelli et al., 2019; Saltelli and Annoni, 2010), part of the problem in the validation of models is that the quality of the accompanying SA is often wanting. Most SA applications still favour the use of a method known as OAT, where the sensitivity of factors is gauged by moving One-factor-At-a-Time (Ferretti et al., 2016; Saltelli et al., 2019). When a sensitivity analysis is run in this fashion, it results in a perfunctory test of the robustness of the model predictions. While different methods exist for sensitivity analysis (see recent reviews in Becker and Saltelli (2015), Borgonovo and Plischke (2016), Iooss and Lemaître (2015), Neumann (2012), Norton (2015), Pianosi et al. (2016), Saltelli et al. (2012), and Wei et al. (2015)), the so-called 'variance-based' methods are considered to be a reference among practitioners. To make an example, when a new method for SA is introduced, its performance is investigated against variance-based measure (see, e.g., Mara et al. (2017)). At present, the most widely used variance-based measures are Sobol' indices (Sobol′, 1993), particularly the Sobol' first-order sensitivity measures $S_j$ and the so-called total sensitivity indices $T_j$ (Homma and Saltelli, 1996). In the following, we take the suggestion from Glen and Isaacs (2012) and for simplicity adopt the symbol $T_j$, rather than $S_{Tj}$ or $S_j^T$, for the total sensitivity indices, although these notations are also commonly found in the literature.

In the next section, we briefly describe how $S_j$ and $T_j$ are defined and computed for the case of independent input factors (Sections 2.1-2.2). Then, we present the set of estimators used (Section 2.3) and define the concepts of economy and explorativity in the estimation procedures for $T_j$ (Saltelli et al., 2010) (Section 2.4). The experimental set up, including the test functions, is outlined in Section 3. Section 4 is dedicated to presenting and discussing our findings, while the general conclusions on the lessons learned are drawn in Section 5.

# 2 Variance-based sensitivity analysis

## 2.1 Variance-based sensitivity measures

For a scalar model output $Y = f(X_1, X_2, \dots, X_k)$, where $X_1$ to $X_k$ are $k$ uncertain factors, the first-order sensitivity index $S_j$ can be written as

$$S_j = \frac{V_{X_j}\left(E_{X_{\sim j}}(Y|X_j)\right)}{V(Y)} \tag{1}$$

where we assume, without any loss of generality and for the case of independent variables, that the factors are uniformly distributed over the $k$-dimensional unit hypercube Ω.

The inner mean in (1) is taken over all-factors-but-$X_j$, (written as $\boldsymbol{X}_{\sim j}$), while the outer variance is taken over factor $X_j$. $V(Y)$ is the unconditional variance of the output variable $Y$.

A short recap of this measure should mention the following without proof (the proof of which can be found in Glen and Isaacs (2012)).

- An efficient way to estimate $S_j$ is to obtain a curve corresponding to $E_{\boldsymbol{X}_{\sim j}}(Y|X_j)$ by smoothing or regressing the scatterplot of $Y$ versus the sorted values of variable $X_j$ and then compute the variance of this curve over $X_j$, as shown in Figure 1.
- The Pearson's correlation ratio squared (Pearson, 1905, 1903), the fraction of the total variability of a response that can be explained by a given set of covariates, commonly indicated as $\eta^2$, coincides with $S_j$.
- When the relationship between $Y$ and $X_j$ is linear, $S_j$ reduces to the squared value of the standardised regression coefficient $\beta^2$, as shown in Figure 1.
- $S_j$ is a first-order term in the variance decomposition of $Y$ (valid when the input factors are independent), which includes terms up to the order $k$, i.e.,

$$1 = \sum_{j=1}^{k} S_j + \sum_{l<j} S_{jl} + \cdots + S_{12\ldots k} \quad (2)$$

- Terms higher than first-order indices are used sparingly in applications due to their multiplicity: a model with $k = 3$ has just three second-order terms, but one with $k = 10$ has as many as forty-five second-order terms. The total number of terms in (2) is $2^k - 1$.
- The meaning of $S_j$ in plain language is 'the fractional reduction in the variance of $Y$ which would be obtained on average if $X_j$ could be fixed'. This is derived from another useful relationship:

$$V_{X_j}\left(E_{\boldsymbol{X}_{\sim j}}(Y|X_j)\right) + E_{X_j}\left(V_{\boldsymbol{X}_{\sim j}}(Y|X_j)\right) = V(Y) \quad (3)$$

  The second term in (3) is the average of all partial variances obtained by fixing $X_j$ to a given value over its uncertainty range. Thus, the first term in (3) is the average reduction. Note that while $V_{\boldsymbol{X}_{\sim j}}(Y|X_j)$ could be greater than $V(Y)$, $E_{X_j}\left(V_{\boldsymbol{X}_{\sim j}}(Y|X_j)\right)$ is always smaller than $V(Y)$ because of (3).

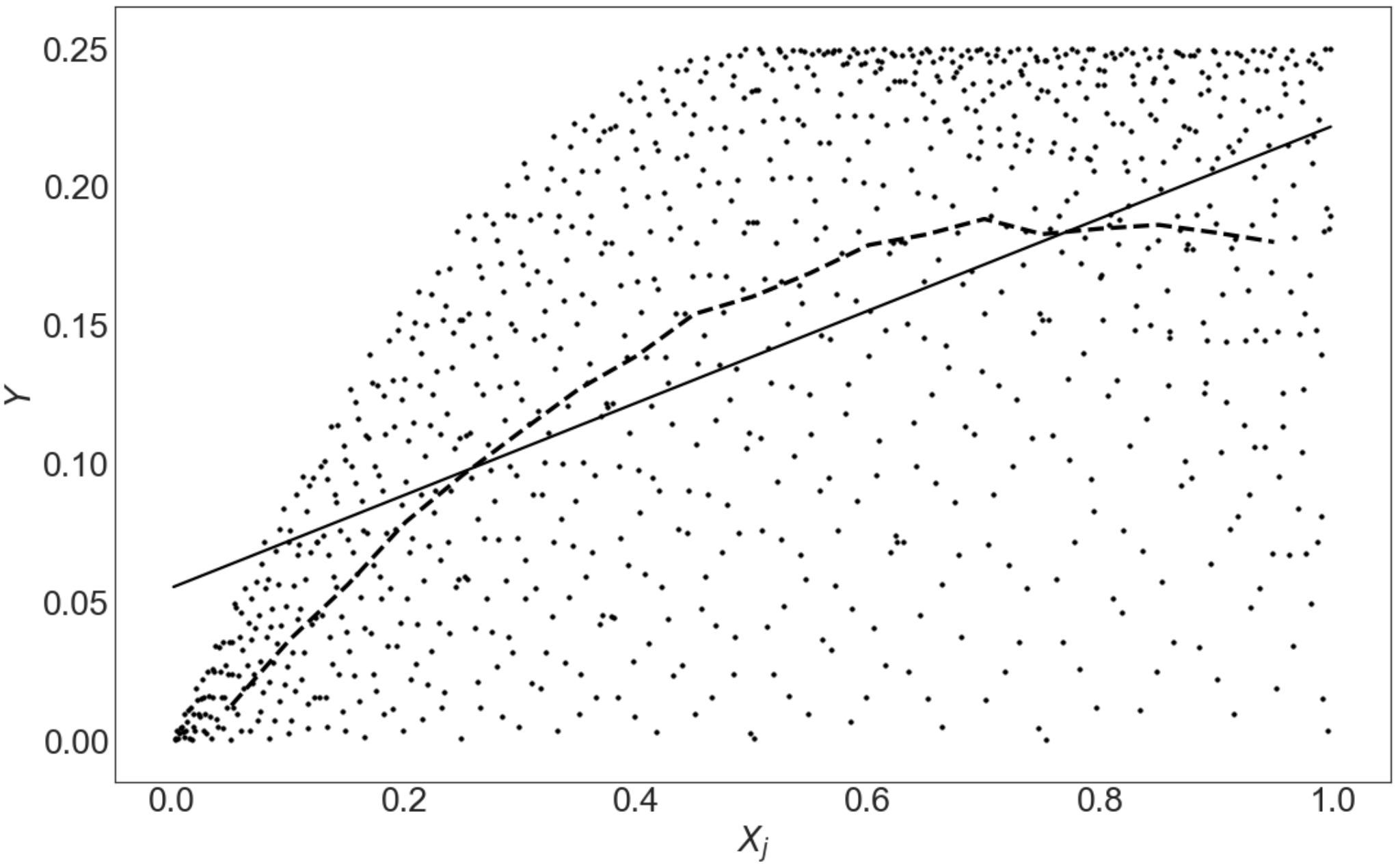

*Figure 1 Sensitivity measures and their relationships in a hypothetical $Y(X_j)$ chart. The dashed line represents the local mean $E_{X_{\sim j}}(Y|X_j)$ of the points in the scatterplot while the straight line corresponds to the standardised regression coefficient $\beta^2$.*

The total-order sensitivity indices $T_j$ can be written as follows:

$$T_j = \frac{E_{\boldsymbol{X}_{\sim j}}\left(V_{X_j}(Y|\boldsymbol{X}_{\sim j})\right)}{V(Y)} \tag{4}$$

The following points are worth recalling for this measure.

- Unlike the case of $S_j$, the smoothing/interpolation of the inner variance is precluded by the impossibility to sort by $\boldsymbol{X}_{\sim j}$ other than by using emulators or Fourier amplitude sensitivity testing, FAST (Saltelli et al., 1999). However, this method requires parametric equations for the search-curve exploring the input space with factor-specific frequencies. Thus, it is more labourious to set up than purely Monte Carlo methods. FAST was the most efficient strategy to compute both $S_j$ and $T_j$ before the work of Saltelli (2002).
- The meaning of $S_j$, in plain language (explicitly descending from Equation (4)), is 'the fraction of variance that would remain on average if one received perfect information on all other factors $X_{\sim j}$'.
- Applying (3) again, one gets

$$E_{\boldsymbol{X}_{\sim j}}\left(V_{X_j}(Y|\boldsymbol{X}_{\sim j})\right) + V_{\boldsymbol{X}_{\sim j}}\left(E_{X_j}(Y|\boldsymbol{X}_{\sim j})\right) = V(Y) \tag{5}$$

Noting that the second term in (5) is the first-order effect on all-but-$X_j$, one derives that the first term in (5), i.e., the numerator in the Equation (4), is the total variance of all terms in decomposition (2) that do include factor $X_j$. For example, for a model with just three factors, one can write

$$1 = S_1 + S_2 + S_3 + S_{12} + S_{13} + S_{23} + S_{123}$$

and

$$T_1 = S_1 + S_{12} + S_{13} + S_{123}$$

- Hence, a parsimonious description of the model $= f(X_1, X_2, \dots, X_k)$ can be obtained by computing all $k$ $S_j$s and all $k$ $T_j$s. This description tells us which factors behave additively $(S_j = T_j)$ and which do not $(S_j < T_j)$. For an additive model, it holds that $S_j = T_j$ for all $j$, and $\sum_{j=1}^{k} S_j = 1$. A limitation of this parsimonious SA is that, in the case of non-negligible interactions, it does not provide information about which factors and which order of interactions are specifically involved.

Computing the couples $S_j, T_j$ can become cumbersome when $Y = f(X_1, X_2, \dots, X_k)$ is computationally time consuming. This could be the case of a mathematical model involving large systems of differential equations, a labourious optimisation programme, natural system simulators involving spatially distributed grid points, and so on. This difficulty is especially relevant for $T_j$, and this is the reason why it is the focus of our contribution.

## 2.2 Sample-based estimation procedures

When estimating a sensitivity measure, two elements are generally demanded: the first is a design, i.e., a strategy to arrange the sample points into the multidimensional space of the input factors; and the second is an estimator, i.e., a formula to compute the selected sensitivity measures. Different authors have suggested different designs and estimators to compute the sensitivity measures (see the contributions reviewed below). In principle, different designs could be tested for a fixed estimator and vice versa, although this is not the most common approach in the literature. In the present work, we have strived to keep the inference and conclusions relative to the design (e.g., in terms of number of sampling matrices) distinct from those relative to the estimator.

The evaluation of sensitivity indices is often based on the integration of Monte Carlo methods. Monte Carlo based procedures for the estimation of $S_j$ have been proposed by Glen and Isaacs (2012), Janon et al. (2014), Lilburne and Tarantola (2009), Mara and Joseph (2008), McKay (1995), Owen (2013), Plischke et al. (2013), Ratto et al. (2007), Saltelli (2002), Saltelli et al. (2010), Sobol′ (1993), and Sobol′ et al. (2007). Some particularly efficient algorithms for the estimation of $S_j$ belong to the class of spectral methods, which may be preferred in case the model has some regularity (Prieur and Tarantola, 2016). These include random balance designs (Tarantola et al., 2006), discrete cosine transformations (Plischke, 2012) and an "Effective Algorithm to compute Sensitivity Indices – EASI" (Plischke, 2010). All of these require a total number of model evaluations that does not depend on the number of factors.

In this paper, we focus solely on $T_j$ and on those estimations based on the actual evaluation of the function at the sampled points without resorting to meta-modelling approaches. We move from the recipe given in Saltelli et al. (2010), which is in turn derived from Saltelli (2002). A quick recap of the main ingredients of this recipe is as follows.

- The computation is based on a quasi Monte Carlo (QMC) method and makes use of quasi-random (QR) points of Sobol' $LP_\tau$ sequences (Sobol, 1976; Sobol′, 1967). QR sequences possess desirable uniformity properties over the unit hypercube Ω. QR numbers are not random: they are designed to optimally fill the unit hypercube in the sense of avoiding inhomogeneous (clustered) points. A useful concept in this respect is that of discrepancy (Kucherenko et al., 2015).
- Given a set of $M$ points inside Ω, their discrepancy is the maximum deviation of the theoretical density ($M$ times the volume of the parallelepiped) over all possible parallelepipeds drawn within the hypercube Ω against the actual density (the number of points in the parallelepiped). Sobol' $LP_\tau$ sequences are designed to be 'low discrepancy' and perform well in existing QR method inter-comparisons (e.g., Sobol′ et al. (2011)).
- In this instance, we use two different sequence generators for Sobol' points: Algorithm 659 (Bratley and Fox, 1988) and SobolSeq16384, distributed by Broda Ltd. (2016) and based on Sobol′ et al. (2011). The different sequences are used in the implementations of our experiment with different programming languages: Python for the former, and Matlab® for the latter.
- A relevant characteristic of $LP_\tau$ sequences is that its uniformity properties deteriorate moving from left to right along a row of the sequence. This means that, for any given $N$, one would expect that the left-most columns of the sample matrix have lower discrepancy than the right-most (Kucherenko et al., 2015).
- The estimation of $T_j$ requires points in Ω that are separated by what is called a 'step in the $X_j$ direction'. In other words, one needs two points in Ω that only differ in the value of factor $X_j$. Note that $S_j$ instead requires steps in the non-$X_j$ direction, e.g., couples of points where all factors but $X_j$ have differing values. As discussed elsewhere (Campolongo et al., 2011), the estimation procedure for $T_j$ resembles the method of Morris. However, $T_j$ is preferred to Morris since the estimation of $T_j$ requires less modelling assumptions and is easier to interpret.
- Given the two $N \times k$ matrices $\boldsymbol{A}$ and $\boldsymbol{B}$**,** we build an additional set of $k$ matrices that we label as $\boldsymbol{A}_{\boldsymbol{B}}^{(j)}$, where column $j$ comes from matrix $\boldsymbol{B}$ and all other $k-1$ columns come from matrix $\boldsymbol{A}$. We indicate that $\boldsymbol{a}_i$ is the $i^{th}$ row of $\boldsymbol{A}$. Likewise, $\boldsymbol{a}_{\boldsymbol{bi}}^{(j)}$ is the $i^{th}$ row of $\boldsymbol{A}_{\boldsymbol{B}}^{(j)}$. Thus, $\boldsymbol{a}_{\boldsymbol{bi}}^{(j)}$ is the $i^{th}$ row of a matrix whose columns come from $\boldsymbol{A}$ except for column $j$ that comes from $\boldsymbol{B}$.

- The model $Y = f(X_1, X_2, \dots, X_k)$ is run for all $f(\boldsymbol{a}_i)$ and $f\left(\boldsymbol{a}_{\boldsymbol{bi}}^{(j)}\right)$ points at a cost of $N(k+1)$ model runs, i.e., $N$ times for the $f(\boldsymbol{a}_i)$ points and $Nk$ times for the $f\left(\boldsymbol{a}_{\boldsymbol{bi}}^{(j)}\right)$ points.
- The numerator in Equation (4), which is needed to compute $T_j$, is obtained from the following estimator (Jansen, 1999):

$$\hat{E}_{\boldsymbol{X}_{\sim j}}\left(V_{X_j}(Y|\boldsymbol{X}_{\sim j})\right) = \frac{1}{2N}\sum_{i=1}^{N}\left[f(\boldsymbol{a}_i) - f\left(\boldsymbol{a}_{\boldsymbol{bi}}^{(j)}\right)\right]^2 \tag{6}$$

- The total variance, the denominator of Equation (4), has been estimated using independent runs, i.e., those corresponding to the rows of matrix $\boldsymbol{A}$.
- Each summand in Equation (6) constitutes an elementary effect fungible for the computation of the total sensitivity index $T_j$.

Sobol′ (2001) noted that this formula (6) was originally proposed by Šaltenis and Dzemyda (1982) (in Russian), and so in the following we shall call it the šaltenis estimator.

In this contribution, we compare the šaltenis estimator with Saltelli's design (Saltelli 2002) to those of Glen and Isaacs (2012), Owen (Iooss et al., 2020), and Lamboni (2018) under a broad set of the test functions.

## 2.3 The examined estimators

The estimators used in Glen and Isaacs (2012) are symmetric (the two base matrices $\boldsymbol{A}$ and $\boldsymbol{B}$ are entrusted the same role and importance) and based on computing the Pearson correlation coefficients between vectors (not to be confused with the Pearson correlation ratio discussed in Section 2.1). This means that for each of the couples of vectors just described (see Table 2), one first computes the correlation coefficients. For example, for the first entry in Table 2, instead of applying the šaltenis estimator (6), one calculates the following:

$$\rho_j = \frac{1}{(N-1)}\sum_{i=1}^{N}\frac{\left(f(\boldsymbol{a}_i)-\langle f(\boldsymbol{a}_i)\rangle\right)\left(f\left(\boldsymbol{a}_{\boldsymbol{bi}}^{(j)}\right)-\langle f\left(\boldsymbol{a}_{\boldsymbol{bi}}^{(j)}\right)\rangle\right)}{\sqrt{V(f(\boldsymbol{a}_i))V\left(f\left(\boldsymbol{a}_{\boldsymbol{bi}}^{(j)}\right)\right)}}, \tag{7}$$

where $\langle f(\boldsymbol{a}_i)\rangle$ is the mean of the $f(\boldsymbol{a}_i)$s over the $N$ runs and $V\left(f(\boldsymbol{a}_i)\right)$ is their variance. This is also the case for $f\left(\boldsymbol{a}_{\boldsymbol{bi}}^{(j)}\right)$. For simplicity, we have not indicated the dependence of $\rho_j$ upon the selected couples of function values $f(\boldsymbol{a}_i)$ and $f\left(\boldsymbol{a}_{\boldsymbol{bi}}^{(j)}\right)$. The best performing estimator according to Glen and Isaacs (2012) – named $D_3$ in their manuscript – has also been used in this study (Equation 8), where the term $c_{d_{-j}}$ is $\frac{1}{2N}\sum_{i=1}^{N}\left(\frac{(f(\boldsymbol{a}_i)-\langle f(\boldsymbol{a}_i)\rangle)\left(f\left(\boldsymbol{a}_{\boldsymbol{bi}}^{(j)}\right)-\langle f\left(\boldsymbol{a}_{\boldsymbol{bi}}^{(j)}\right)\rangle\right)}{\sqrt{V(f(\boldsymbol{a}_i))V\left(f\left(\boldsymbol{a}_{\boldsymbol{bi}}^{(j)}\right)\right)}}+\frac{(f(\boldsymbol{b}_i)-\langle f(\boldsymbol{b}_i)\rangle)\left(f\left(\boldsymbol{b}_{\boldsymbol{ai}}^{(j)}\right)-\langle f\left(\boldsymbol{b}_{\boldsymbol{ai}}^{(j)}\right)\rangle\right)}{\sqrt{V(f(\boldsymbol{b}_i))V\left(f\left(\boldsymbol{b}_{\boldsymbol{ai}}^{(j)}\right)\right)}}\right)$ and the other terms are detailed in the Appendix (Table A1).

$$\hat{T}_j = 1 - c_{d_{-j}} + p_j \frac{c_{a_j}}{1 - c_{a_j} c_{a_{-j}}} \tag{8}$$

Additionally, Glen and Isaacs (2012) note that supposedly uncorrelated vectors, such as $f(\boldsymbol{a}_i)$ and $f(\boldsymbol{b}_i)$ or $f\left(\boldsymbol{a}_{\boldsymbol{bi}}^{(j)}\right)$ and $f\left(\boldsymbol{b}_{\boldsymbol{ai}}^{(j)}\right)$ (where $\boldsymbol{b}_{\boldsymbol{ai}}^{(j)}$ is the $i^{th}$ row of matrix $\boldsymbol{B}_{\boldsymbol{A}}^{(j)}$), may be affected by spurious correlations for finite values of $N$. We say 'supposedly uncorrelated' since no columns are shared between $f(\boldsymbol{a}_i)$ and $f(\boldsymbol{b}_i)$, nor is this the case for $f\left(\boldsymbol{a}_{\boldsymbol{bi}}^{(j)}\right)$ and $f\left(\boldsymbol{b}_{\boldsymbol{ai}}^{(j)}\right)$. These spurious correlations are explicitly computed in Glen and Isaacs (2012) and then used as correction terms in the computation of the sensitivity indices.

The main advantage of the symmetric design proposed over the asymmetric design of Saltelli et al. (2010) is that the coordinates of the base sample appear disproportionately with respect to the other coordinates in Saltelli (2002) while this is not the case with Glen and Isaacs (2012).

To assess whether the use of multiple matrices can be beneficial, we tested the Owen estimator (Iooss et al., 2020) as Equation (9) (where $\boldsymbol{c}_{\boldsymbol{bi}}^{(j)}$ is the $i^{th}$ row of matrix $\boldsymbol{C}_{\boldsymbol{B}}^{(j)}$). According to the author (Owen, 2013), three-base matrices estimators offer better accuracy than two-base matrices ones.

$$\hat{E}_{\boldsymbol{X}_{\sim j}}\left(V_{X_j}(Y|\boldsymbol{X}_{\sim j})\right) = V(Y) - \frac{1}{N}\sum_{i=1}^{N}\left[f(\boldsymbol{b}_i) - f\left(\boldsymbol{c}_{\boldsymbol{bi}}^{(j)}\right)\right]\left[f\left(\boldsymbol{b}_{\boldsymbol{ai}}^{(j)}\right) - f(\boldsymbol{a}_i)\right] \tag{9}$$

The same case is made by Lamboni (2018) as regards the use of estimators having even more base matrices as per Equation (10)

$$\hat{E}_{\boldsymbol{X}_{\sim j}}\left(V_{X_j}(Y|\boldsymbol{X}_{\sim j})\right) = \frac{n-1}{Nn^2}\sum_{i=1}^{N}\sum_{m=1}^{n}\left[\sum_{q=1,q\neq m}^{n}\frac{1}{n-1}\left[f(\boldsymbol{h}_{m,i}) - f\left(\boldsymbol{h}_{m-q,i}^{(j)}\right)\right]\right]^2 \tag{10}$$

where $\boldsymbol{h}_{m,i}$ is a generic row $i$ of a base matrix and $\boldsymbol{h}_{m-q,i}^{(j)}$ the same row but for coordinate $j$, as well as the Saltelli et al. (2010) asymmetric design of the šaltenis estimator for a variable number of matrices.

## 2.4 Economy and explorativity of the design of estimators

As discussed in Saltelli et al. (2010), there is a natural trade-off between the economy of a design (how many elementary effects we can obtain with a given number of runs, as in Equation 11) and how well the design fills or explores the input-factor space, i.e., in our case, the unit hypercube Ω. These aspects are taken here as measures of the quality of different estimators. In this contribution, economy ($e$) and explorativity ($\chi$) are defined in the context of the calculation of $T_j$.

The trade-off between the economy ($e$) and explorativity ($\chi$) of the hypercube comes from the fact that one should strive to use any given point more than once to make a computation efficient. However, the more a given point is re-used, the less the $k$-dimensional space of the input is explored. In practice, when using $n$ matrices $\boldsymbol{A}$, $\boldsymbol{B}$, $\boldsymbol{C}$, etc. each of column length $N$, as well as hybrid matrices such as $\boldsymbol{A}_{\boldsymbol{B}}^{(i)}$, the following situation occurs: the $nN$ points corresponding to the $n$ matrices $\boldsymbol{A}$, $\boldsymbol{B}$, $\boldsymbol{C}$, etc. will all contain original coordinates while all the hybrid matrices such as $\boldsymbol{A}_{\boldsymbol{B}}^{(i)}$ reuse coordinates. The points are in general different, i.e., no point in $\boldsymbol{A}_{\boldsymbol{B}}^{(i)}$ coincides with points of either $\boldsymbol{A}$ or $\boldsymbol{B}$ (except for the very few rows in the beginning of Sobol' $LP_\tau$ sequences, which include repeated coordinates) while, for example, $\boldsymbol{A}$ and $\boldsymbol{A}_{\boldsymbol{B}}^{(i)}$ share $k-1$ columns and $N(k-1)$ coordinates in total. We shall show in a moment how the number of non-repeated coordinates for a fixed total number of runs diminishes by increasing the number of matrices $n$.

We know from the previous subsection that the estimator described in Equation (6) yields $N$ elementary effects per factor, i.e., $Nk$ differences $f(\boldsymbol{a}_i) - f\left(\boldsymbol{a}_{\boldsymbol{b}i}^{(j)}\right)$ are used in Equation (6) at the cost of $N(k+1)$ runs of the model. The economy $e$ of the design is thus the following:

$$e = \frac{E_T}{N_T} = \frac{Nk}{N(k+1)} = \frac{k}{k+1} \tag{11}$$

which is less than one.

From now on we shall call this design 'asymmetric' due to the different roles entrusted to matrices $\boldsymbol{A}$ and $\boldsymbol{B}$: the coordinates of $\boldsymbol{A}$ are used more than those of $\boldsymbol{B}$.

Saltelli et al. (2010) also tried to use a larger number $n > 2$ of base matrices. The idea was that with, e.g., $n = 3$ matrices, $\boldsymbol{B}$, $\boldsymbol{C}$, and always using $\boldsymbol{A}$ as the base sample matrix, one would have $\binom{n}{2} = \binom{3}{2} = 3$ ways of generating elementary effects. In addition to couples of function values $f(\boldsymbol{a}_i), f\left(\boldsymbol{a}_{\boldsymbol{b}i}^{(j)}\right)$, one can also use the couples $f(\boldsymbol{a}_i), f\left(\boldsymbol{a}_{\boldsymbol{c}i}^{(j)}\right)$ and $f\left(\boldsymbol{a}_{\boldsymbol{b}i}^{(j)}\right), f\left(\boldsymbol{a}_{\boldsymbol{c}i}^{(j)}\right)$. All these couples are in fact only one step $X_j$ apart. This design produces $3Nk$ elementary effects at the cost of $N(1+2k)$ runs for an economy of ${3k}/{(1+2k)}$, which is greater than one. By computing all functional values for the three matrices (and not just for $\boldsymbol{A}$), one computes $3N$ functional values $f(\boldsymbol{a}_i), f(\boldsymbol{b}_i)$, and $f(\boldsymbol{c}_i)$.

The next functional values corresponding to all possible mixed matrices are $f\left(\boldsymbol{a}_{\boldsymbol{b}i}^{(j)}\right)$ and $f\left(\boldsymbol{a}_{\boldsymbol{c}i}^{(j)}\right)$ but also $f\left(\boldsymbol{b}_{\boldsymbol{a}i}^{(j)}\right)$, $f\left(\boldsymbol{b}_{\boldsymbol{c}i}^{(j)}\right)$, $f\left(\boldsymbol{c}_{\boldsymbol{a}i}^{(j)}\right)$, and $f\left(\boldsymbol{c}_{\boldsymbol{b}i}^{(j)}\right)$. An additional set of $6Nk$ function values for a total of $3N(1+2k)$ runs has been generated. Each of the three matrices $\boldsymbol{A}$, $\boldsymbol{B}$ and $\boldsymbol{C}$ can be used to compute $2Nk$ effects, as shown in the first six rows of Table 2 below. $3Nk$ additional effects can be obtained by mixing the hybrid matrices (last three rows in Table 2).

| ***Table 2*** *Couplings leading to elementary effects,* i.e.*, couples of function values fungible for the computation of* $T_j$ *in the case of* $n = 3$ *matrices* | |
|---|---|
| $f(\boldsymbol{a}_i)$ | $f\left(\boldsymbol{a}_{\boldsymbol{bi}}^{(j)}\right)$ |
| $f(\boldsymbol{a}_i)$ | $f\left(\boldsymbol{a}_{\boldsymbol{ci}}^{(j)}\right)$ |
| $f(\boldsymbol{b}_i)$ | $f\left(\boldsymbol{b}_{\boldsymbol{ai}}^{(j)}\right)$ |
| $f(\boldsymbol{b}_i)$ | $f\left(\boldsymbol{b}_{\boldsymbol{ci}}^{(j)}\right)$ |
| $f(\boldsymbol{c}_i)$ | $f\left(\boldsymbol{c}_{\boldsymbol{ai}}^{(j)}\right)$ |
| $f(\boldsymbol{c}_i)$ | $f\left(\boldsymbol{c}_{\boldsymbol{bi}}^{(j)}\right)$ |
| $f\left(\boldsymbol{a}_{\boldsymbol{bi}}^{(j)}\right)$ | $f\left(\boldsymbol{a}_{\boldsymbol{ci}}^{(j)}\right)$ |
| $f\left(\boldsymbol{b}_{\boldsymbol{ai}}^{(j)}\right)$ | $f\left(\boldsymbol{b}_{\boldsymbol{ci}}^{(j)}\right)$ |
| $f\left(\boldsymbol{c}_{\boldsymbol{ai}}^{(j)}\right)$ | $f\left(\boldsymbol{c}_{\boldsymbol{bi}}^{(j)}\right)$ |

This gives a total of $9Nk$ effects for the case of $n = 3$ matrices for an economy $e = {}^{9Nk}\!/_{3N(1+2k)} = {}^{3k}\!/_{(1+2k)}$.

How can this be extended to a design with a generic number of matrices? Given $n$ matrices, there are $\binom{n}{2}$ pairwise combinations; and for each of the $2k$ matrices that are produced, there is twice the number of factors (since for each couple of matrices, such as $\boldsymbol{A}$ and $\boldsymbol{B}$, we shall have to consider both matrices $\boldsymbol{B}_{\boldsymbol{A}}^{(j)}$and $\boldsymbol{A}_{\boldsymbol{B}}^{(j)}$). Since each matrix is composed of $N$ runs, $N_T$ will be $N\left(n + 2k\binom{n}{2}\right) = N\big(n + kn(n-1)\big) = nN\big(1 + k(n-1)\big)$. With similar considerations, one derives that the number of effects $E_T$ will be $Nkn(n-1) + 3Nk\binom{n}{3} = Nkn(n-1) + \frac{1}{2}Nkn(n-1)(n-2) = \frac{1}{2}Nkn^2(n-1)$.

In summary, we have

$$N_T = nN\big(1 + k(n-1)\big) \tag{12}$$

$$E_T = \frac{1}{2}Nkn^2(n-1) \tag{13}$$

and the resulting economy is defined in Eq. (14), whereby the value of $e$ tends to ${}^{n}\!/_{2}$ for a large $n$ or/and a large $k$.

$$e = \frac{E_T}{N_T} = \frac{\frac{1}{2}Nkn^2(n-1)}{nN(1+k(n-1))} = \frac{kn(n-1)}{2(1+k(n-1))} \quad (14)$$

Note that the same development made for $T_j$ could be replicated for $S_j$, although first-order indices are not in the scope of this manuscript.

Different arrangements can be explored to calculate $T_j$ to have the couples of points differing for the $j^{th}$ coordinate only. These settings can be compared against how many coordinates are used out of the maximum number of $N_T k$coordinates available (explorativity, $\chi$). The lower $\chi$ is, the less explorative the design, although the design's economy ($e$) may be increased in these settings. Possible arrangements are detailed as follows. We recall from our legend (Table 1) that economy here is relative to the sole computation of the elementary effects useful to estimate $T_j$ for all the $k$ factors.

**Couples**. The $N_T$ points are arbitrarily arranged in ${}^{N_T}/_2$ couples. Each couple needs just *k* coordinates for one point and only one extra coordinate for the companion point. The points of each couple differ for one of the *j* coordinates. To produce this arrangement, one needs to generate ${}^{N_T(k+1)}/_2$ coordinates. Thus, the fraction $\chi$ of coordinates relative to the maximum $N_T k$ is

$$\chi = \frac{N_T(k+1)/2}{N_T k} = \frac{k+1}{2k} \quad (15)$$

which tends to ½ as *k* increases.

**Stars.** $\frac{N_T}{k+1}$ points are initially generated in the hypercube, which is the core of the star, from which each of the available *k* dimensions is explored in turn. In this way, each star is made of *k+1* points and needs *k+k* coordinates: *k* for the centre point of the star and one for each of its *k* rays. Thus, the fraction $\chi$ is now the follows:

$$\chi = \frac{2N_T k}{N_T(k+1)k} = \frac{2}{k+1} \quad (16)$$

which decreases as k increases.

**Winding stairs (one trajectory)**. In a winding-stair design, the hypercube is explored using a curve whereby each coordinate is increased in turn. This design needs *k* coordinates to generate the first point and $N_T - 1$ additional coordinates for the remaining points

$$\chi = \frac{N_T+k-1}{N_T k} \quad (17)$$

which generally tends to *1/k* as *N>>k.* This changes if one uses more than one trajectory. For example, if one uses trajectories of length *k+1*, the explorativity becomes identical to that of 'stars' above.

**Saltelli et al. (2010) asymmetric.** The design needs one base matrix $\boldsymbol{A}$ and $k$ additional matrices $\boldsymbol{A}_{\boldsymbol{B}}^{(1)}, \boldsymbol{A}_{\boldsymbol{B}}^{(2)} .. \boldsymbol{A}_{\boldsymbol{B}}^{(k)}$ with column length *N*. This corresponds to a total of $N_T = N(k+1)$ points for a total of $Nk(k+1)$*)* coordinates. However, one only needs a total of *2Nk* coordinates, *Nk* for each of the two matrices ***A*** and ***B***. Hence,

$$\chi = \frac{2Nk}{Nk(k+1)} = \frac{2}{k+1} \tag{18}$$

which is the same as the 'stars' above. This design is one of the most widely used standards in the literature against which we will be benchmark the performance of the assessed estimators.

**Saltelli (2002) symmetric.** In this design, one makes use of both sets $\boldsymbol{A}_{\boldsymbol{B}}^{(1)}, \boldsymbol{A}_{\boldsymbol{B}}^{(2)} .. \boldsymbol{A}_{\boldsymbol{B}}^{(k)}$ and $\boldsymbol{B}_{\boldsymbol{A}}^{(1)}, \boldsymbol{B}_{\boldsymbol{A}}^{(2)} .. \boldsymbol{B}_{\boldsymbol{A}}^{(k)}$ for a total of $N_T = 2N(k+1)$ points, which in principle would correspond to a total of $N_T = 2Nk(k+1)$ coordinates. However, only *2Nk* coordinates are needed, *Nk* for each of the two matrices ***A*** and ***B***. Hence,

$$\chi = \frac{2Nk}{2Nk(k+1)} = \frac{1}{k+1} \tag{19}$$

**Glen and Isaacs (2012)** (estimator D3, symmetric). This is the same as **Saltelli 2002, symmetric** above.

**Owen (Iooss et al., 2020)** (Equation 9). Only the matrices $\boldsymbol{A}$, $\boldsymbol{B}$, $\boldsymbol{B}_{\boldsymbol{A}}^{(j)}$, and $\boldsymbol{C}_{\boldsymbol{B}}^{(j)}$ are used in this estimator for a total of cost of $N_T = 2N(k+1)$. Equation 19 is based only on one summation; hence, it counts only for one effect. Thus, $E_T = Nk$ and $e = \frac{Nk}{2N(k+1)} = \frac{k}{2(k+1)}$. The number of coordinates used is $2Nk(k+1)$, of which only $3Nk$ are non-repeated.

$$\chi = \frac{3kN}{2kN(k+1)} = \frac{3}{2(k+1)} \tag{20}$$

**Generalisation of the symmetric case.** The design now includes *n* base matrices ***A, B, …X***, where ***X*** is the n[th] matrix, plus a total of two times $\binom{n}{2}k$ additional hybrid matrices of the type $\boldsymbol{A}_{\boldsymbol{B}}^{(1)}, \boldsymbol{A}_{\boldsymbol{B}}^{(2)} .. \boldsymbol{A}_{\boldsymbol{B}}^{(k)}$ and $\boldsymbol{B}_{\boldsymbol{A}}^{(1)}, \boldsymbol{B}_{\boldsymbol{A}}^{(2)} .. \boldsymbol{B}_{\boldsymbol{A}}^{(k)}$, where the binomial corresponds to the possible number of couplings of two matrices. Thus, the total number of matrices is $n + 2\binom{n}{2}k = n + n(n-1)k = n\big(1 + k(n-1)\big)$. The total number of points is $N_T = nN\big(1 + k(n-1)\big)$, corresponding to a maximum number of coordinates $nNk\big(1 + k(n-1)\big)$. Since the number of coordinates used in this design is just those of the base matrices, i.e., *nNk*, the fraction $\chi$ is

$$\chi = \frac{nNk}{nNk(1+k(n-1))} = \frac{1}{1+k(n-1)} \tag{21}$$

which decreases as *n* increases and reduces to (15) for *n=2*.

**Lamboni (2018)** (Equation 10) The total cost is the same $N_T = nN(1 + k(n-1))$ as above for multiple matrices. $E_T = Nkn(n-1)$ effects are computed in this case since the $3Nk\binom{n}{3}$ differences among couples of mixed matrices are not accounted for in this estimator. Hence, $e = \frac{Nkn(n-1)}{nN(1+k(n-1))} = \frac{k(n-1)}{(1+k(n-1))}$ while the explorativity $\chi$ is the same as in the generalisation of the symmetric case showed in Equation 21.

The economy and explorativity of the different designs are compared in Figure 2

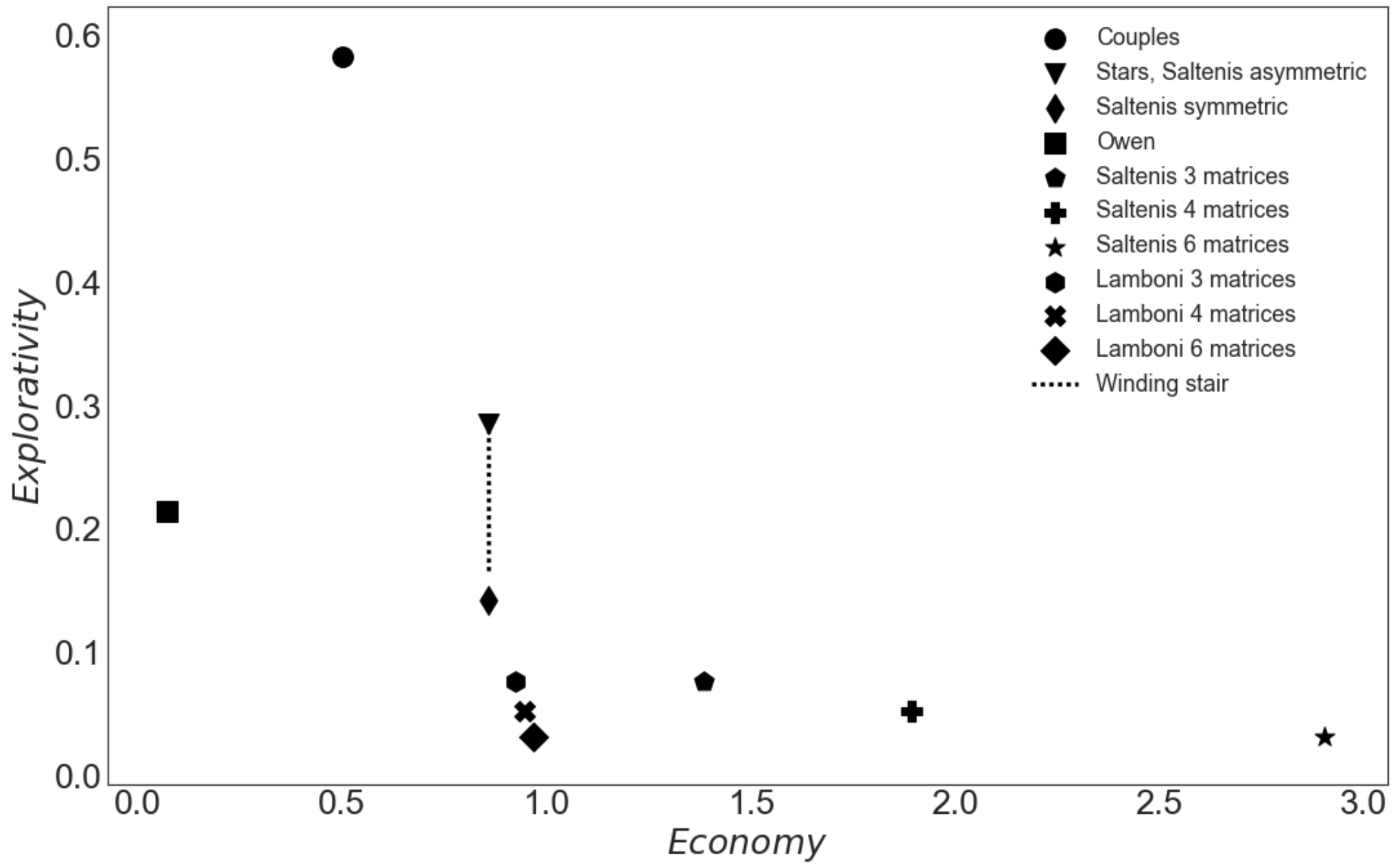

*Figure 2 Explorativity $\chi$ vs economy e for the different designs in the case of $k = 6$ factors. The winding stair interval is defined by the explorativity $\chi$ on the minimum number of points ($N_T = k + 1, \chi = {}^{2}/_{(k+1)}$) and the asymptotic case ($N_T \gg k, \chi = {}^{1}/_{k}$).*

The overall cost of the sensitivity analysis in terms of $N_T$ and the number of factors $k$ are typically known to the modeller prior to the analysis. Hence, different couples of $n$ and $N$ can be chosen to meet the target $N_T$ value. To maximise $E_T$, one would set $n$ as high as possible. In terms of discrepancy $(D)$ and $\chi$, one would rather have lower values of $n$ to have less repeated coordinates. The lower the fraction of repeated coordinates (the higher $\chi$) is, the better the space-filling properties of the design, and hence $D$ is lower. Therefore, $D$ and $\chi$ have an inverse relation.

Note that the relation between the discrepancy and error is not simple. A given $D$ can be perfect for a smooth function and inadequate for a jigsaw-shaped one. With fixed $k$ and $N_T$, there is an inverse quadratic relation between $n$ and $N$, as shown in Equation (12), which describes the trade-off between $e$ and $\chi$.

Let us examine the case where $k = 6$ and the cost $N_T \approx 500$ runs (Table 3). Since $N$ needs to be a power of two in quasi-random number sequences based on Sobol' $LP_\tau$ sequences, the value of $N_T$ may deviate from 500. $D$ has been estimated using the computational method provided by Jäckel (2002) and rounded to two digits (Table 3).

***Table 3 -*** *Possible values of $n$ and $N$ corresponding to a model with $k = 6$ factors, an affordable $N_T \cong 500$ and discrepancy $D$. $N_T$ has been adjusted to have a power of 2 for $N$, as requested in the QMC simulations based on Sobol' $LP_\tau$ sequences.*

| $N$ | $n$ | $N_T$ | $E_T$ | $nN$ | $D$ | $\chi$ |
|---|---|---|---|---|---|---|
| 64 | 2 (asymmetric) | 448 | 384 | 128 | 0.0065 | 0.27 |
| 32 | 2 (symmetric) | 448 | 384 | 64 | 0.0076 | 0.13 |
| 16 | 3 (symmetric) | 624 | 864 | 48 | 0.013 | 0.077 |
| 8 | 4 (symmetric) | 608 | 1152 | 32 | 0.020 | 0.053 |
| 4 | 5 (symmetric) | 500 | 1200 | 20 | 0.032 | 0.04 |
| 2 | 7 (symmetric) | 518 | 1764 | 14 | 0.053 | 0.027 |
| 1 | 10 (symmetric) | 550 | 2700 | 10 | 0.11 | 0.018 |

The last row has the highest number of effects using the smallest number of random points –, i.e., just one row of a single QMC matrix in six dimensions. The opposite applies to the first row since it uses as many as 64 rows from two QMC matrices, but it gives the fewest effects. In other words, the first row is the most explorative while the last is the most economical in terms of the number of effects per run.

Using $n > 2$ (Saltelli et al., 2010) resulted in poorer convergence with respect to the case $n = 2$. According to that paper, the šaltenis estimator in conjunction with $n = 2$ was the best available sample-based practice. Contrasting findings have been reported by Lamboni (2018), according to whom the optimal number of matrices may be different from two depending on the function evaluated.

We also tested whether a variable explorativity $\chi$ across factors could improve the accuracy of the estimate. This experiment consisted of allocating more runs to the factors having the highest standard deviation of the elementary effects after an initial warm up. The number of residual runs is attributed according to the importance obtained at that given point in the simulation. The computational details of this experiment are described in section 4.

# 3 Experimental set up - Test cases

## 3.1 Test Functions

Saltelli et al. (2010) and Glen and Isaacs (2012) base their analysis on a single function, namely, the widely used G function, which is defined as follows:

$$G(X_1, X_2, \dots, X_k, a_1, a_2, \dots, a_k) = \prod_{j=1}^{k} g_j, \quad (22)$$

with

$$g_j = \frac{|4X_j - 2| + a_j}{1 + a_j} \quad (23)$$

With this test function, one can modulate the importance of a factor $X_j$ via the associated constant $a_j$, as shown in Table 4.

***Table 4** – Factors' importance in the function G dependent on the constant $a_j$*

| $a_j$ | **Importance of $X_j$** |
|---|---|
| 0 - 0.99 | Very high |
| 0.99 – 9.9 | High |
| 9.9 – 99 | Non-important |
| $> 99$ | Non-significant |

Although this function can be attributed to Davis and Rabinowitz (1984), it was further developed by Saltelli and Sobol′ (1995) and is known among practitioners as Sobol's $G$ function. It is reduced to the function used in Davis and Rabinowitz (1984) when all $a_j$ = 0.

A six-dimensional version of the $G$ function with coefficients $a_j = \{0\ 0.5\ 3\ 9\ 99\ 99\}$ is used here as in Glen and Isaacs (2012), Saltelli et al. (2010), and function A2 below (Equation 25). To test the effectiveness of the estimators with a wider typology of functions, we have used the taxonomy suggested by Kucherenko et al. (2011) in Equations (24 –30).

$$A1: f(X) = \sum_{j=1}^{k} (-1)^j \prod_{l=1}^{j} x_j \quad (24)$$

$$A2: f(X) = \prod_{j=1}^{k} \frac{|4x_j - 2| + a_j}{1 + a_j} \quad (25)$$

$$B1: f(X) = \prod_{j=1}^{k} \frac{k - x_j}{k - 0.5} \quad (26)$$

$$B2: f(X) = (1+\frac{1}{k})^k \prod_{j=1}^{k} \sqrt[k]{x_j} \quad (27)$$

$$B3: f(X) = \prod_{j=1}^{k} \frac{|4x_j-2|+a_j}{1+a_j} \quad (28)$$

$$C1: f(X) = \prod_{j=1}^{k} |4x_j-2| \quad (29)$$

$$C2: f(X) = 2^k \prod_{j=1}^{k} x_j \quad (30)$$

The analytical values for the sensitivity indices are available (Kucherenko et al., 2011; Saltelli et al., 2010) from the GitHub repository: https://github.com/Confareneoclassico/New_estimator_algorithm. Following the taxonomy of Kucherenko et al. (2011), the functions of group A are the easiest for SA, with only a few important factors and low cross-factor interactions. In class B functions, all the factors are important, but the cross-factor interactions are low. Class C functions are the most difficult to treat for SA with non-negligible interactions across all important factors.

When the coefficients $a_j$ of function $G$ are all equal and large, one is dealing with a B-type function (B3, Equation 28, for which all $a_j$=6.42). By contrast, the case of $a_j$ null coefficients (Davis and Rabinowitz, 1984) would correspond to a C-type function (function C1 above, Equation 29).

## 3.2 Computational arrangements

The Python code used for the computations reported in the present work is available from the following *GitHub* repository: https://github.com/Confareneoclassico/New_estimator_algorithm. A second Matlab® code was used in a separate set of computations limited to one test function (A2): https://github.com/Confareneoclassico/Variance_SA_estimators_designs_explorativity_economy/tree/master/MatlabCode. The agreement of the independent computations coded and run by separate co-authors (SLP and FF) is offered as internal validation of the results presented in this paper.

Each test comparison is repeated 50 times to ensure reproducibility and reduce the stochastic variation in the results. Some of the experiments were also run 500 times to ensure stability. However, no major difference was observed between 50 and 500 repetitions: https://github.com/Confareneoclassico/Variance_SA_estimators_designs_explorativity_economy/tree/master/Extra_material. Each repetition uses an equal number $N$ of quasi-random rows from the Sobol' matrix with the input factors $x_j$ uniformly distributed in (0,1). The total cost of the analysis is kept consistent across methods.

For each of the 50 repetitions, the randomisation procedure is based on the column permutations of the QMC matrix. The first thirty-six columns (that correspond to $6k$ since $n = 6$ is the largest number of multiple-matrix design tests) of the Sobol' sequence are scrambled in each repetition. The $k$ left-most ones are attributed to matrix **A,** the following $k$ to matrix **B** and so forth depending on the number of matrices assessed in the estimator.

As customarily done in QMC computations using Sobol' sequences, the column dimension $N$ of each matrix is rounded to the nearest power of two: each power of two corresponds to a 'full scan' of the hyperspace for each block of the sequence. Different blocks of size $2^p$ ($p = 2, 3, \ldots 14$) have been tested for selected functions (classes A, B and C) against $N_T$.

Following Saltelli et al. (2010), the simulation results are presented in terms of the mean absolute error (MAE) versus the total cost of the analysis, where the MAE is defined as follows:

$$MAE = \frac{1}{50}\sum_{r=1}^{50}\left(\frac{\sum_{j=1}^{k}|\hat{T}_j - T_j|}{k}\right)_r \quad (31)$$

where $T_j$ is the analytical value of the total sensitivity index, and $\hat{T}_j$ is its estimated value. In other words, the total error over all factors is considered for the difference between the numerical estimate of the index (averaged over all available elementary effects) and its analytic value. The results are plotted on a decimal logarithmic scale.

# 4 Results and discussion

The best-performing estimator suggested by Glen and Isaacs (2012), named by them as estimator D3, is compared against šaltenis with the Saltelli asymmetric design (from now on the *Šaltenis asymmetric estimator* for short) (Figure 3). These results have also been confirmed by the Matlab® implementation.

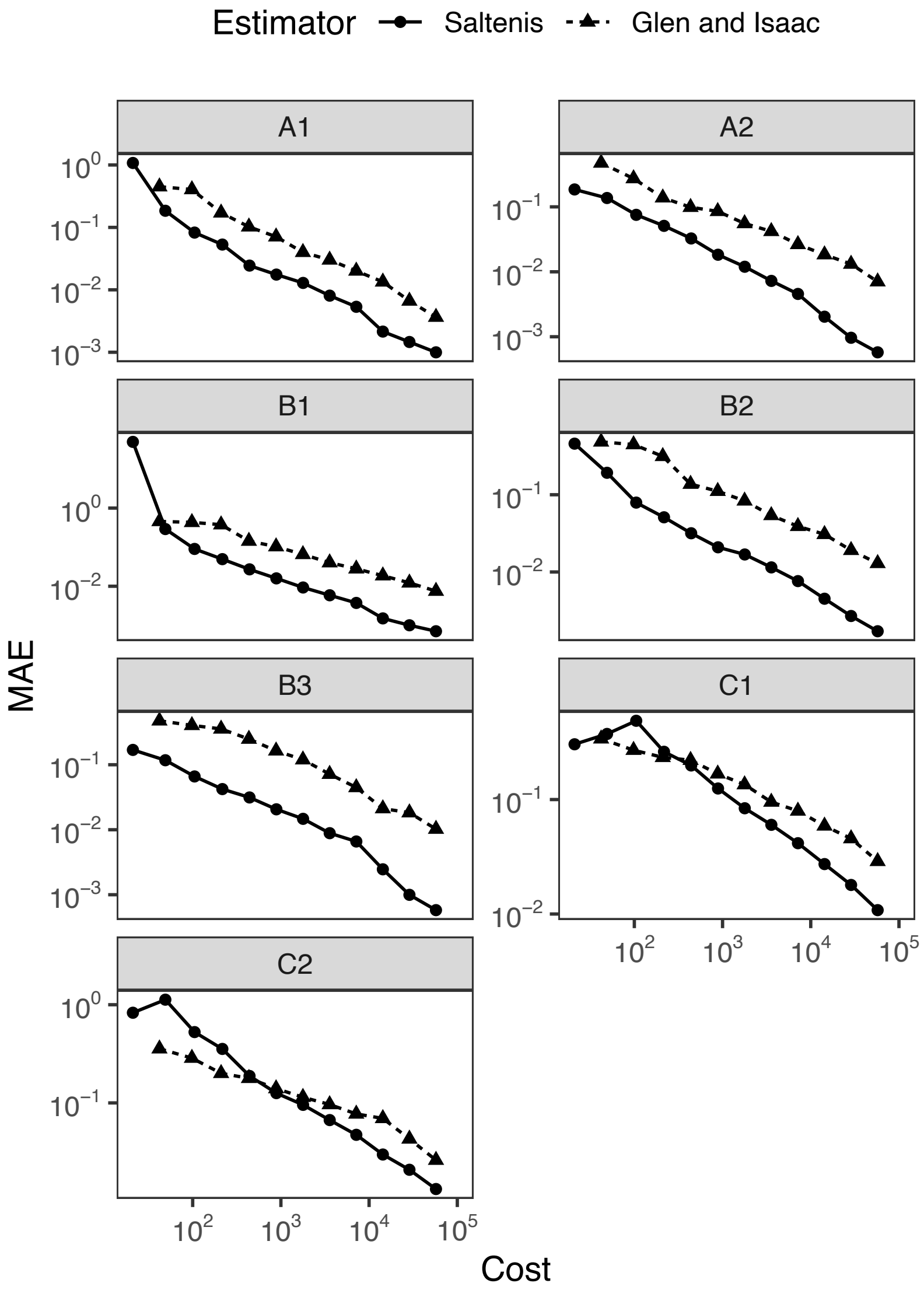

*Figure 3 MAE vs cost ($N_T$) on a logarithmic scale for the Šaltenis asymmetric estimator (circle, continuous line)and the Glen-Isaacs estimator (triangle, dashed line). Functions: A1, A2, B1, B2, B3, C1, and C2 (Eq. 24-30). Python implementation. Glen-Isaacs cannot achieve a very low cost due to its higher cost evaluating the same number of rows evaluation with respect to Šaltenis.*

As previously discussed, computing $S_j$ requires couples of points where all factors but $X_j$ have differing values. The logic of correcting these sets of points for spurious correlations is that we are considering vectors such as $f(\boldsymbol{a}_i)$ and $f\left(\boldsymbol{b}_{\boldsymbol{a}i}^{(j)}\right)$, where $i = 1,2, \dots N$, when computing the correlation $\rho_j$ for the sensitivity index $S_j$. If any of these columns is spuriously correlated in the two matrices because of the finite value of $N$, then this spurious correlation should be removed from $\rho_j$ as described in Glen and Isaacs (2012).

These authors suggest that a similar correction is useful for computing $T_j$. However, the *Šaltenis asymmetric estimator* for $T_j$ shows better convergence properties for type A and B functions and marginal better convergence properties for type C functions. This can be explained by considering that the computation of $T_j$ requires vectors such as $f(\boldsymbol{a}_i)$ and $f\left(\boldsymbol{a}_{\boldsymbol{b}i}^{(j)}\right)$, where $i = 1,2, \dots N,$ where now all columns but $j$ are identical in the two vectors, and the differing column $j$ is the one under investigation. There are no chances of strong spurious correlations in this case. Looking back to Table 3 or Figure 2 tells us that one should not expect an improvement when changing from the asymmetric to the symmetric case for $n = 2$ because we obtain the same number of effects at the cost of halving the explorativity of the design.

The Owen estimator (Iooss et al., 2020) has higher explorativity, as seen in section 2.4. This estimator makes different use of the set of base matrices $\boldsymbol{A}, \boldsymbol{B},$ and $\boldsymbol{C}$ to improve the computation of the elementary effects. However, one can see in Figure 4, that the *Šaltenis asymmetric estimator* systematically outperforms this estimator.

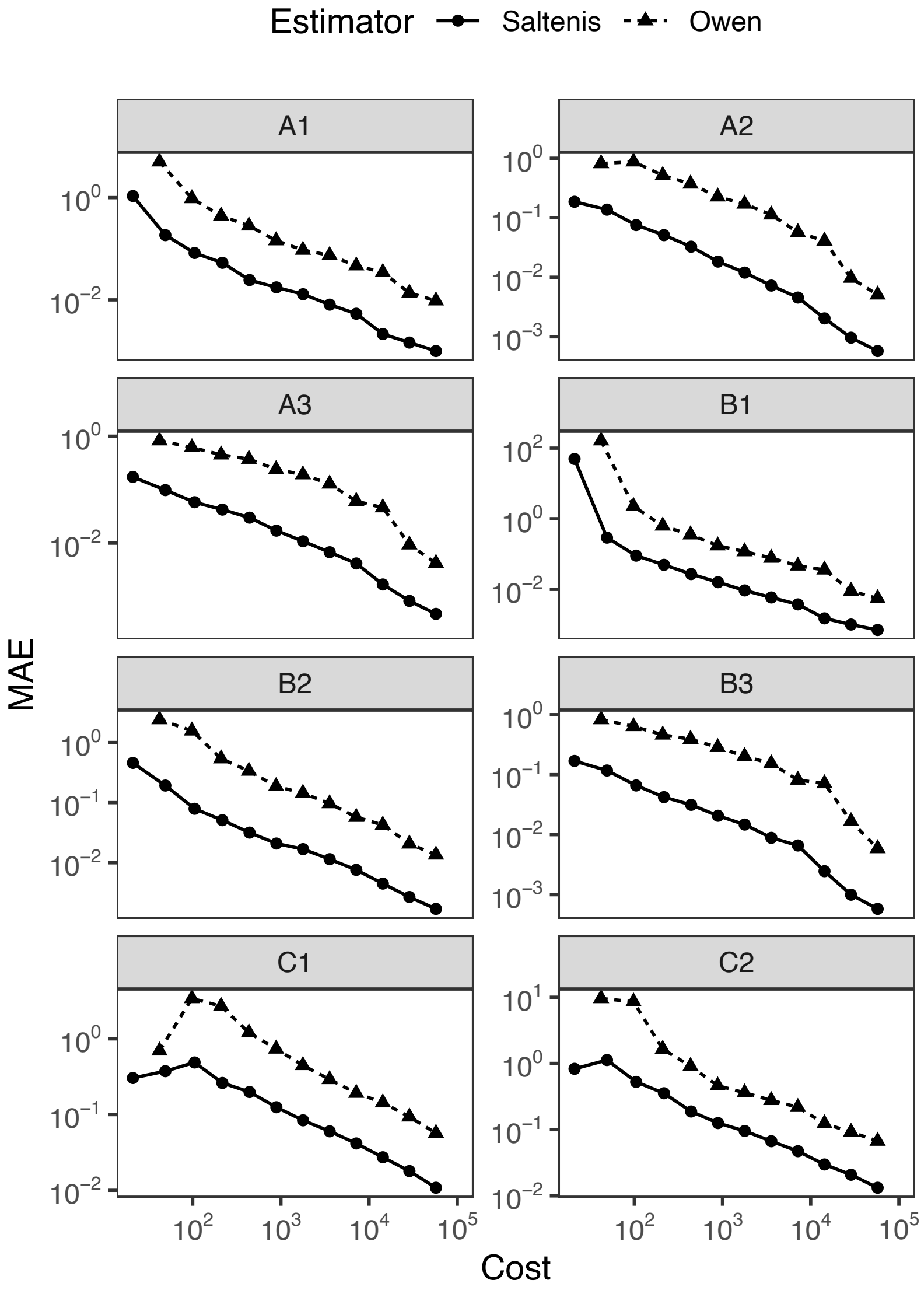

*Figure 4 - MAE vs cost ($N_T$) on a logarithmic scale for the Šaltenis asymmetric estimator (circle, continuous line) and the Owen estimator (triangle, dashed line). Functions: A1, A2, B1, B2, B3, C1, and C2 (Eq. 24-30). Python implementation. Owen cannot achieve a very low cost due to its higher cost evaluating the same number of rows with respect to Šaltenis.*

Moving to the case of multiple-matrix-based designs, one would have hoped that moving to $n > 2$ could improve $e$ because the decrease in $\chi$ is offset by an increased number of effects as per Table 3, but this does not appear to be the case (Figure 5). The asymmetric design based on just two matrices is still the best option to compute the total sensitivity indices $T_j$, even when compared against the symmetric design for the *Šaltenis asymmetric estimator* and those based on multiple matrices (Figure 5). It would thus appear that $\chi$ is more important than $e$: the increased number of effects is outperformed by the decreased number of original coordinates. While a slight mismatch in the costs exists due to the different number of sample matrices, the performance gap is significant to the point that the estimator lagging behind is not capable of catching up with the front runner, even when doubling the sample size in most of the cases assessed. The differences are slightly decreasing from the A- and B-functions to the C-functions.

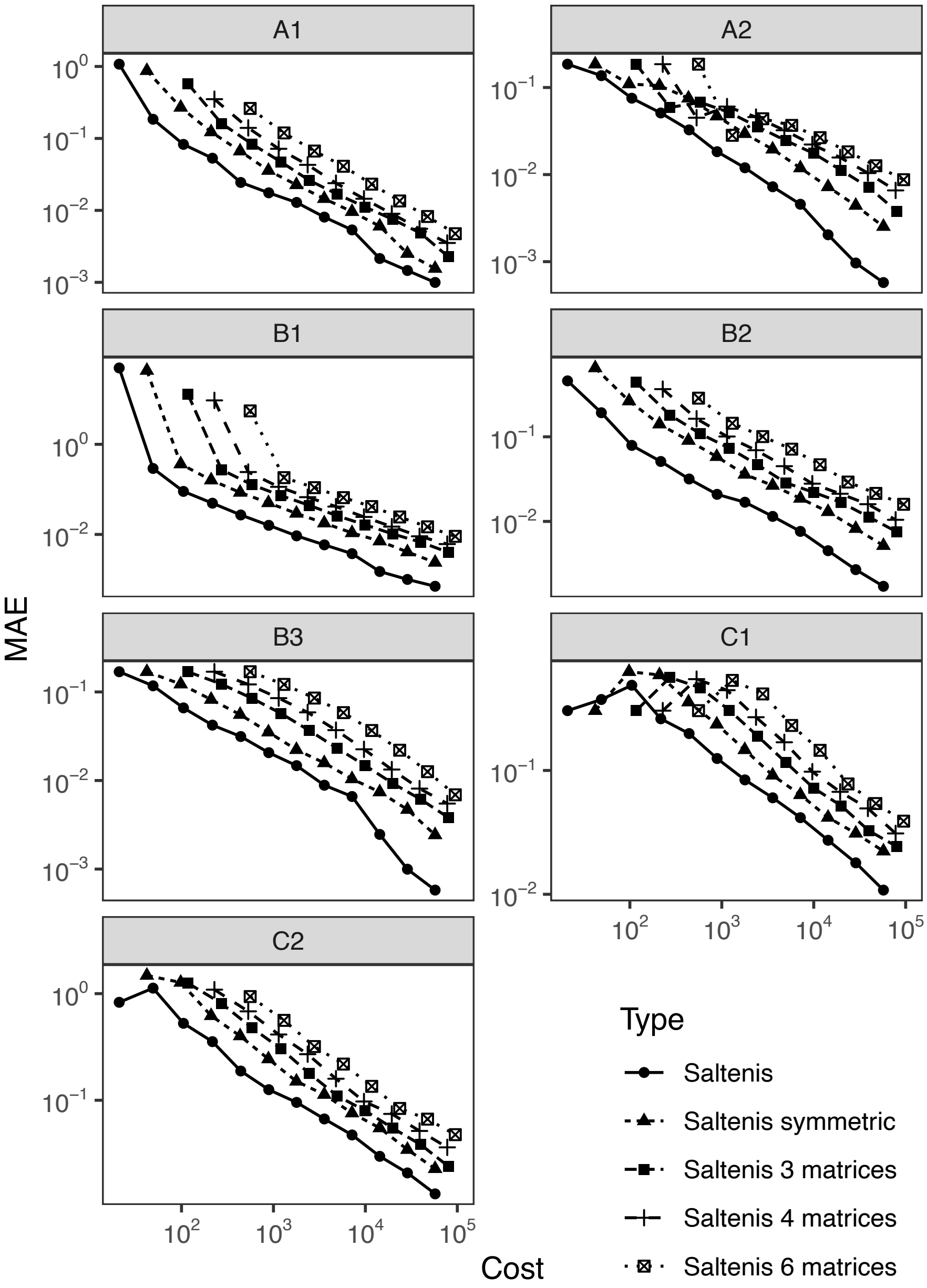

*Figure 5 MAE vs cost ($N_T$) on a logarithmic scale for the Šaltenis asymmetric estimator (circle, continuous line), two-matrix-based symmetric estimator (triangle, dashed line), three-matrix-based estimator (square, dot-dashed-dotted line), four-matrix-based estimator (cross, dash-dotted line) and six-matrix-based estimator (empty square, cross-dotted line). Functions: A1, A2, B1, B2, B3, C1, and C2 (Eq. 24-30). Python implementation.*

The results of the comparison with the Lamboni estimator are examined in Figure 6. Note that the two-matrix symmetric design of the *Šaltenis* estimator corresponds to Lamboni's for this number of matrices. The lower distance of the Lamboni estimator from the frontrunner (the *Šaltenis asymmetric estimator*) confirms that this design outperforms the multiple-matrix-based design for the šaltenis estimator (comparing Figure 5 and Figure 6). However, it is still beaten by the *Šaltenis asymmetric estimator*. The result is also confirmed when the difference with the analytic error is measured as the root-mean squared error (not shown).

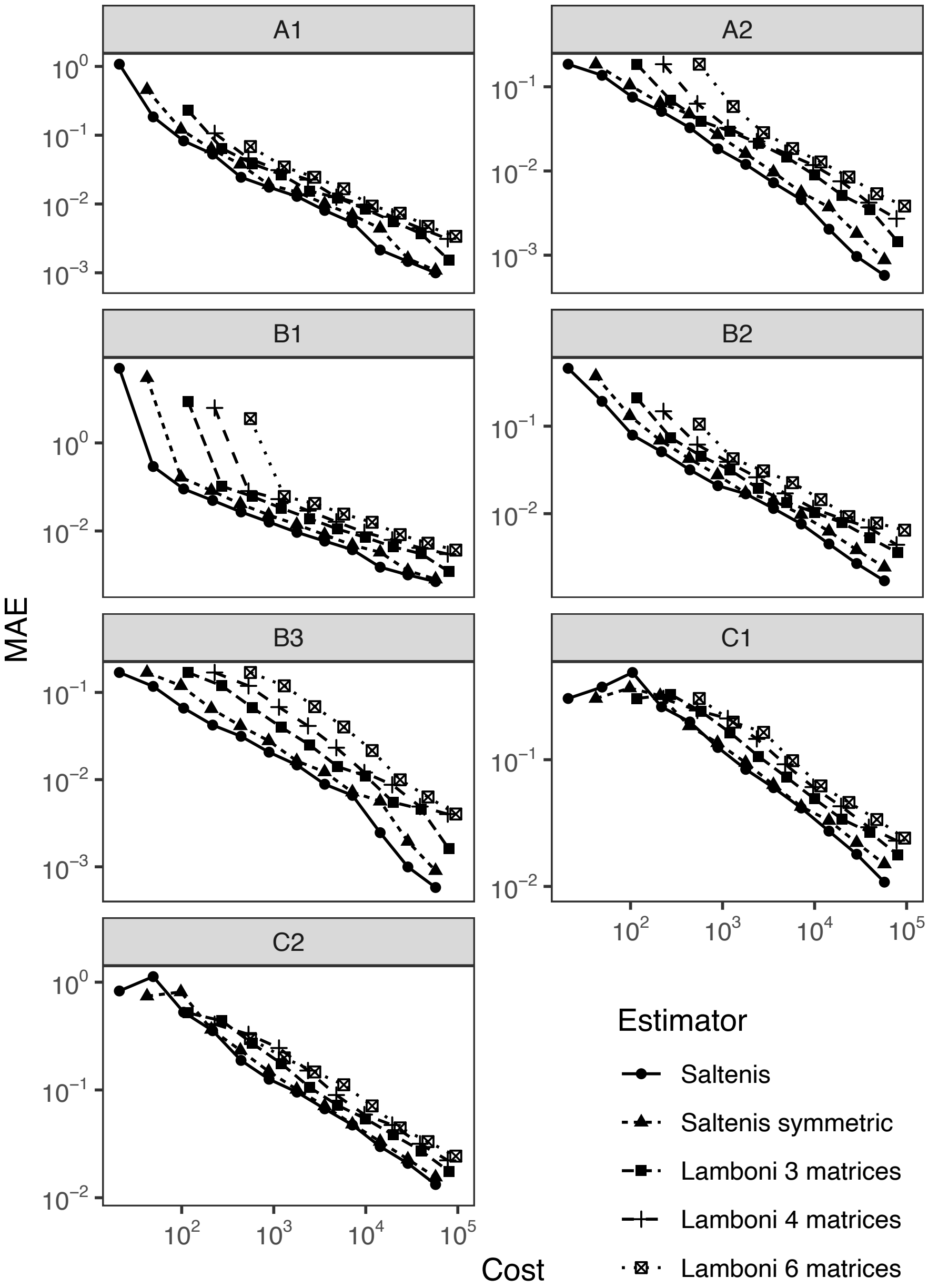

*Figure 6 - MAE vs cost ($N_T$) on a logarithmic scale for the Šaltenis asymmetric estimator (circle, continuous line), two-matrix-based symmetric estimator (triangle, dashed line), Lamboni three-matrix-based estimator (square, dot-dashed-dotted line), Lamboni four-matrix-based estimator (cross, dash-dotted line) and. Lamboni six-matrix-based estimator (empty square, cross-dotted line). Functions: A1, A2, B1, B2, B3, C1, and C2 (Eq. 24-30). Python implementation.*

As shown above, the trade-off between $e$ and $\chi$ demonstrates that $n > 2$ is not a convenient design choice. Another way to look at these results is to assess them in terms of stars, which are computationally equivalent to the Saltelli asymmetric design (Saltelli et al., 2010), as seen in section 2.4. The basic design is the one where each star is made of *k+1* points using *2k* coordinates. To increase $e$, one must increase the number of points in the stars, although this results in decreasing $\chi$ since one uses more coordinates of the core. The cases where the number of matrices is greater than 2 fall into this class. This approach led to worse results. In other words, increasing $e$ does not seem to pay off.

We have also tried to compute $T_j$ using matrix $\boldsymbol{A}$ alone, i.e., instead of computing $T_j$ from $f(\boldsymbol{a}_i)$ and $f\left(\boldsymbol{a}_{\boldsymbol{b}i}^{(j)}\right)$, we used $f(\boldsymbol{a}_i)$ and $f\left(\boldsymbol{a}_{i,(i+1)}^{(j)}\right)$. In this approach, when the last N-th row is reached, one uses $f(\boldsymbol{a}_N)$ and $f\left(\boldsymbol{a}_{N,1}^{(j)}\right)$ for $T_j$, i.e., the system closes on itself. However, this approach did not lead to improvements.

Another sampling procedure we have tested to improve the Šaltenis asymmetric estimator consisted of variably investing the computational budget by improving $T_j$'s estimation for the subset of the most important factors while devoting less computational resources to the least important ones for each subsequent model execution. In this adaptive sampling strategy, the choices of the factor to estimate are made using increasing blocks of power of two to fully take advantage of the properties of the low-discrepancy Sobol' sequence.

The number of design parameters of the adaptive sampling strategy is reduced to a modicum. Let one assume that one has $N_T = (k + 1)2^p$ model runs available.

- The algorithm is run as per the Šaltenis asymmetric estimator (Saltelli et al., 2010) to 'warm up' to a sample size $N$=$2^{p+2-k}$ at a cost of $(k + 1)2^{p+2-k}$.
- The $k$ factors are then ordered in decreasing order of the standard deviation of the elementary effect $std_{ee}^{j}$.
- At every following block of rows $N$=$2^{s+p+2-k}$ (with $s$ in the range $1, k - 1$), it is decided whether the computation of the elementary effect can be stopped at the $(k - s)^{th}$ factor per order of decreasing importance, thus saving runs.
- The condition upon which this decision is made is the ratio between $std_{ee}^{j}$-s. It is assumed that $std_{ee}^{j}$ would be reduced by a factor of $\sqrt{2}$ by doubling the sample size.

- Hence, the main assumption of the computation is that if ${std_{ee}^{k-s-1}}/{\sqrt{2}} > std_{ee}^{k-s}$, this latter factor (and all those having lower $std_{ee}^{j}$s) can be removed from the calculation in the following block.

The computational details are available from the dedicated Jupyter notebook. The results are here presented for test functions A1 and A2, for which $T_j$ differs across parameters. Another type of function G has also been tested in this experiment, where A3 ($a_j = \{1\ 2\ 4\ 8\ 16\ 32\}$) corresponds to various degrees of importance across parameters.

In Figure 7, one can appreciate how our method outperforms the *Šaltenis asymmetric estimator* by up to a factor of two for functions A2 and A3. In this context, the importance of input variables on the output uncertainty can be easily disentangled due to the difference in magnitude across factors' $T_j$. This is the setting of a typical real-world model, where the importance of the input factors on the output uncertainty obeys the Pareto principle (Pareto, 1906) with few factors responsible for most of the output variance. However, the case where the sensitivity indices of the input factors are closer in magnitude is more challenging. This adaptive sampling strategy does not outperform the *Šaltenis asymmetric estimator* in the case of function A1 (Figure 7).

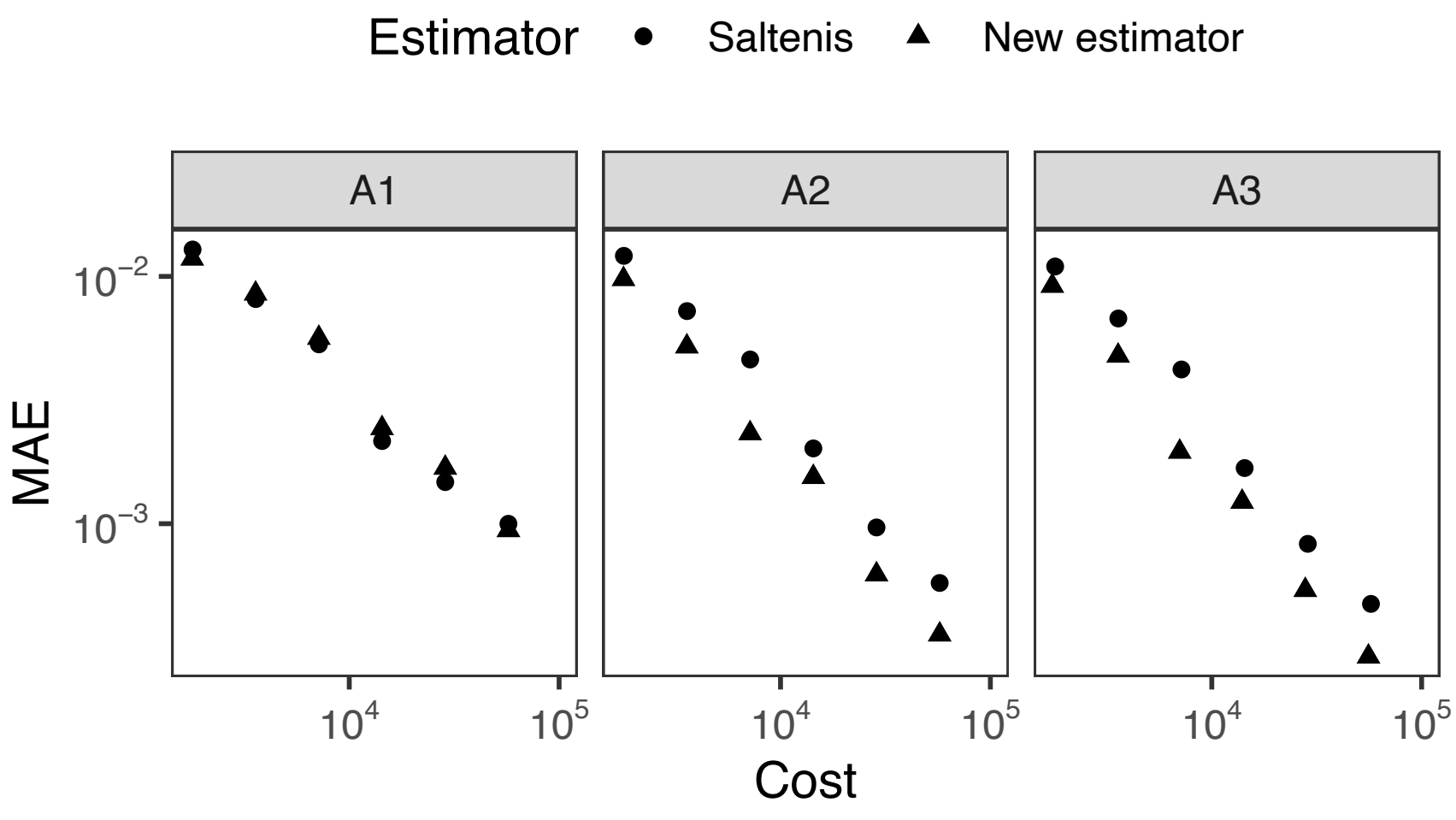

*Figure 7 - MAE vs cost ($N_T$) on a logarithmic scale for the Šaltenis asymmetric estimator (circle) and the proposed adaptive sampling strategy (triangle). Each point corresponds to the MAE reported at full cost $N_T = (k+1)2^p$. Functions: A1 and A2 (Eqs. 24-25, respectively) and A3.*

## 5 Conclusions

Taking the works of Glen and Isaacs (2012), Lamboni (2018), and Saltelli et al. (2010) as our points of departure, we have explored different estimators to improve the computation of the total-effect index $T_j$ for independent factors using a taxonomy of test functions proposed by Kucherenko (2011).

We have seen that the estimator of Glen and Isaacs (2012) is outperformed by Šaltenis and Dzemyda (1982) and the Saltelli asymmetric design (Saltelli et al., 2010). Furthermore, we did not observe improvements in the computational results by extending the symmetric matrix arrangement to values $n > 2$. The larger number of effects obtained with $n > 2$ does not compensate for the loss of explorativity, as is also evidenced by our discrepancy calculation. The increase in economy by using more matrices is offset by the loss of explorativity due to the higher share of repeated coordinates.

To increase the explorativity, one would need to rely on a 'stars' design with centres having less than $k$ rays, which decreases the economy. The latter approach has led to an improvement in the setting of factors receiving a number of estimates proportional to their importance. However, this comes at the cost of introducing an extra design parameter.

# 6 Acknowledgments

Elmar Plischke from Technische Universität Clausthal, Guillaume Damblin from CEA, Sergei Kucherenko from Imperial College of London, Stefano Tarantola and Thierry Mara from the Joint Research Centre of the European Commission, and three anonymous reviewers, offered useful comments and suggestions. The remaining errors are uniquely due to the authors. This work was partially funded by a grant from the Peder Sather Centre for Advances Studies of the University of Berkeley "Mainstreaming Sensitivity Analysis and Uncertainty Auditing", awarded in June 2017. Arnald Puy has worked on this paper on a Marie Sklodowska-Curie Global Fellowship, grant number 792178.

# A Appendix

***Table A1*** *– Terms composing the Glen & Isaacs (2012) estimator D3*

| Term | Explicit formula |
|---|---|
| $c_{d_{-j}}$ | $\frac{1}{2N}\sum_{i=1}^{N}\left(\frac{(f(\boldsymbol{a}_i)-\langle f(\boldsymbol{a}_i)\rangle)\left(f\left(\boldsymbol{a}_{\boldsymbol{bi}}^{(j)}\right)-\left\langle f\left(\boldsymbol{a}_{\boldsymbol{bi}}^{(j)}\right)\right\rangle\right)}{\sqrt{V(f(\boldsymbol{a}_i))V\left(f\left(\boldsymbol{a}_{\boldsymbol{bi}}^{(j)}\right)\right)}}+\frac{(f(\boldsymbol{b}_i)-\langle f(\boldsymbol{b}_i)\rangle)\left(f\left(\boldsymbol{b}_{\boldsymbol{ai}}^{(j)}\right)-\left\langle f\left(\boldsymbol{b}_{\boldsymbol{ai}}^{(j)}\right)\right\rangle\right)}{\sqrt{V(f(\boldsymbol{b}_i))V\left(f\left(\boldsymbol{b}_{\boldsymbol{ai}}^{(j)}\right)\right)}}\right)$ |
| $c_{d_j}$ | $\frac{1}{2N}\sum_{i=1}^{N}\left(\frac{(f(\boldsymbol{b}_i)-\langle f(\boldsymbol{b}_i)\rangle)\left(f\left(\boldsymbol{a}_{\boldsymbol{bi}}^{(j)}\right)-\left\langle f\left(\boldsymbol{a}_{\boldsymbol{bi}}^{(j)}\right)\right\rangle\right)}{\sqrt{V(f(\boldsymbol{b}_i))V\left(f\left(\boldsymbol{a}_{\boldsymbol{bi}}^{(j)}\right)\right)}}+\frac{(f(\boldsymbol{a}_i)-\langle f(\boldsymbol{a}_i)\rangle)\left(f\left(\boldsymbol{b}_{\boldsymbol{ai}}^{(j)}\right)-\left\langle f\left(\boldsymbol{b}_{\boldsymbol{ai}}^{(j)}\right)\right\rangle\right)}{\sqrt{V(f(\boldsymbol{a}_i))V\left(f\left(\boldsymbol{b}_{\boldsymbol{ai}}^{(j)}\right)\right)}}\right)$ |
| $p_j$ | $\frac{1}{2N}\sum_{i=1}^{N}\left(\frac{\left(f(\boldsymbol{a}_i)-\langle f(\boldsymbol{a}_i)\rangle\right)\left(f\left(\boldsymbol{b}_i^{(j)}\right)-\langle f\left(\boldsymbol{b}_i^{(j)}\right)\rangle\right)}{\sqrt{V\left(f(\boldsymbol{a}_i)\right)V\left(f\left(\boldsymbol{b}_i^{(j)}\right)\right)}} + \frac{\left(f\left(\boldsymbol{a}_{\boldsymbol{bi}}^{(j)}\right)-\langle f\left(\boldsymbol{a}_{\boldsymbol{bi}}^{(j)}\right)\rangle\right)\left(f\left(\boldsymbol{b}_{\boldsymbol{ai}}^{(j)}\right)-\langle f\left(\boldsymbol{b}_{\boldsymbol{ai}}^{(j)}\right)\rangle\right)}{\sqrt{V\left(f\left(\boldsymbol{a}_{\boldsymbol{bi}}^{(j)}\right)\right)V\left(f\left(\boldsymbol{b}_{\boldsymbol{ai}}^{(j)}\right)\right)}}\right)$ |
| $c_{a\,j}$ | $\frac{c_{d_j}-p_j c_{d_{-j}}}{1-p_j^2}$ |
| $c_{a-j}$ | $\frac{c_{d_{-j}}-p_j c_{d_j}}{1-p_j^2}$ |